 \documentstyle[preprint,aps,pra,tighten]{revtex}
\begin{document}
\draft
{\tighten
\title{A Simple Method for Calculating Quantum Effects\\
on the Temperature Dependence of Bimolecular Reaction Rates:\\
An Application to CH$_4$ + H $\to$ CH$_3$ + H$_2$}}
\author{David Z. Goodson,$^{*,}$\protect\cite{smu}
Dustin W. Roelse,\cite{smu} Wan-Ting Chiang,\protect\cite{smu}
Steven M. Valone,\protect\cite{lanl} and\\ J. D. Doll\protect\cite{brown}}
\address{Contribution from the
Department of Chemistry, Southern Methodist University, Dallas,
Texas 75275, Materials Science and Technology Division, MST-7,
Los Alamos National Laboratory,
Los Alamos, New Mexico 87545, and Department of Chemistry, Brown
University, Providence,\\ Rhode Island  02912}
\maketitle
\begin{abstract}
The temperature dependence of the rate of the reaction
CH$_4$+H$\,\to\,$CH$_3$+H$_2$ is studied
using classical collision theory with a temperature-dependent
effective potential derived from a path integral analysis.
Analytical expressions are obtained for the effective
potential and for the rate constant.  The rate constant
expressions use a temperature-dependent activation energy.
They give better agreement with the available experimental
results than do previous empirical fits.  Since all but
one of the parameters in the present expressions are obtained
from theory, rather than by fitting to experimental reaction
rates, the expressions can be expected to be more dependable
than purely empirical expressions at temperatures above
2000 K or below 350 K, where experimental results are
not available.
\end{abstract}
\pacs{}

\narrowtext

\section{Introduction}

The validity of the Arrhenius expression for the rate constant
of a bimolecular reaction, $k=A\exp(-E_a/RT)$, where $E_a$ is
the activation energy and $R$ is the gas constant, has
long been the subject of controversy \cite{johnston}.  $E_a$
is generally assumed to be independent of temperature, and
any deviation from linear behavior in the plot of $\log k$
vs. $1/T$ is attributed to temperature dependence in $A$.
Classical collision theory \cite{collisiontheory} suggests only a
weak temperature dependence of the preexponential factor, of the
form $A\propto T^{1/2}$, while theoretical arguments based on
transition-state theory (TST) with corrections for quantum mechanical
tunneling \cite{johnston,frost} can predict a significantly different
dependence.

We consider here the reaction
\begin{equation}
{\rm CH}_4 + {\rm H} \rightarrow {\rm CH}_3 + {\rm H}_2.
\label{reaction}
\end{equation}
There exists an extensive literature of theoretical
and experimental studies
\cite{clark,shaw,tsang,schatz,joseph,truhlar,truong,walker,smp,baulch,rabinowitz}
for this reaction on account of its importance as an
elementary reaction in hydrocarbon pyrolysis.
The experimental results span the temperature range from 372 K
up to almost 2000 K.  TST studies
\cite{clark,shaw,tsang,schatz,joseph,truhlar,truong}
have predicted a distinct upward curvature in the Arrhenius plot,
and an analysis by Shaw \cite{shaw}
of experimental results through the year 1978 generally
supported this prediction.  Subsequently, Sepehrad {\it et al.}
\cite{smp} concluded that in fact a linear Arrhenius plot was
more consistent with the data, after omitting some apparently
unreliable earlier results and adding new results of their
own.  More recent analyses by Baulch {\it et al.} \cite{baulch}
and by Rabinowitz {\it et al.} \cite{rabinowitz} discerned
curvature with $A\propto T^3$ and $A\propto T^{2.11}$,
respectively.

We propose an alternative theoretical approach to this problem,
based not on transition-state theory but on classical
collision theory with a temperature-dependent effective
potential energy function.  This approach is based on
a path-integral analysis developed by
Feynman \cite{feynman0,feynman1} for problems in quantum
statistical mechanics.  He noted that the equation for
the statistical density matrix is formally identical to that for
the kernel that expresses the time dependence of the
wavefunction of a quantum mechanical particle over a time
interval that is taken to be negative and imaginary.  Thus,
calculations in statistical mechanics can be carried out
using the path integral techniques of quantum dynamics.
This idea can be used to derive an effective potential
$V_{\rm eff}$ for the chemical reaction such that a
classical mechanical calculation of the reaction rate with
the effective potential is approximately equivalent to a
quantum mechanical treatment using the actual potential $V$
\cite{doll}.
This simple approach was used previously to describe the diffusion
of H on a Cu surface \cite{valone}, and the results were
later found to agree with those from an elaborate
reaction-path variational TST calculation with semiclassical
adiabatic tunneling corrections \cite{lauderdale}.

The path integral analysis transforms $V$ into a
{\it temperature-dependent} function $V_{\rm eff}(T)$, from
which we obtain a temperature-dependent activation energy.
Thus, our model for the chemical reaction is classical collision
theory but with a temperature-dependent $E_a$.
We will include no temperature dependence in $A$ other than the
classical $T^{1/2}$ factor.  We will introduce a slight modification
into the analysis so that the path-integral result, which is
derived as a perturbation theory about the high-$T$ limit,
can be smoothly interpolated to the correct low-$T$ result.
For $V$ we will use the empirical hydrocarbon potential of
Brenner \cite{brenner}, which is constructed from Morse-type
functions with modifications to take into account non-local
effects.

\section{Method}

To calculate the partition function from the statistical
density matrix it is sufficient to
consider only paths that return to their starting point.
The path integrals that need to be evaluated are very
difficult on account of the large number of degrees of
freedom needed to describe the many possible paths.
However, if $\hbar/k_BT$ is small, where $k_B$ is the
Boltzmann constant, then one can derive
a simple approximate expression for the partition function,
\cite{feynman0,feynman1},
\begin{equation}
Z=\left({mk_BT\over 2\pi\hbar^2}\right)^{1/2}
\int e^{-\beta V_{\rm eff}(x)}dx,
\label{partitionfn}
\end{equation}
where $\beta=1/k_BT$ and
\begin{equation}
V_{\rm eff}(x)= {1\over\sqrt{2\pi}\sigma}
\int_{-\infty}^\infty V(x+y)e^{-y^2/2\sigma^2}dy,
\label{veff}
\end{equation}
\begin{equation}
\sigma^2=\hbar^2/12mk_BT.
\label{sigma}
\end{equation}
The significance of this result is the fact that
Eq.~(\ref{partitionfn}) has exactly the form of the {\it classical}
partition function except that $V$ is replaced by the
effective potential $V_{\rm eff}$, which is just a Gaussian
average of $V$ with a temperature-dependent standard deviation
$\sigma$.  Equation (\ref{partitionfn}) is for a system with only
one degree of freedom, but the extension to an arbitrary number
of degrees of freedom is straightforward.

It has been suggested \cite{doll} that classical dynamics
on the potential $V_{\rm eff}$ could be used to simulate
quantum dynamics on the true potential $V$.  This would not
be valid for the dynamics of a single particle, but it is a
reasonable hypothesis for a statistical ensemble of particles
as in a molecular dynamics simulation of a chemical reaction.
In fact, it can be shown \cite{doll} that when $V_{\rm eff}$
is used in place of $V$ in TST one obtains the standard
Wigner tunneling correction.

This approach provides an appealing qualitative model for
quantum effects.  At a minimum of $V$, averaging over
neighboring points according to Eq.~(\ref{veff}) will increase
the potential.  This accounts for the fact that the minimum of
$V$ is in practice inaccessible to the system on account of the
impossibility of localizing a quantum mechanical particle.  In
effect, the averaging provides a zero-point energy correction.
At a maximum, the averaging reduces $V$.  In effect, this is a
tunneling correction.  At a saddle point of $V$, averaging
over a given coordinate will reduce the potential if $V''$
is negative and increase it if $V''$ is positive.  Elsewhere,
$V$ can be approximated as a linear function, in which case
the averaging in Eq.~(\ref{veff}) will have little effect.  At
high $T$ the quantum effects will be small, because if the average
kinetic energy is large then it is only rarely that a particle will
be close enough to the potential surface to sense the
difference between $V$ and $V_{\rm eff}$.  Accordingly, the
standard deviation given by Eq.~(\ref{sigma}) goes to zero in
the limit of infinite $T$.

At low $T$ the approximations used to derive Eq.~(\ref{veff})
can be expected to lead to
a significant error, since the system will spend much of its
time in the quantum regions of $V$.  However, a minor
modification of the theory will ensure that it give the
correct low-temperature limit.  At $T=0$ K the system will
be at rest at the nearest local minimum of $V$.  The energy
of the system will be the value of $V$ plus a zero-point
energy correction.  A minimum of $V$ in principle corresponds
to a stable chemical species, for which the zero-point energy
can be determined empirically from analysis of the vibrational
spectrum.  Thus, we can replace $T$ in Eq.~(\ref{sigma}) with
\begin{equation}
T_{\rm eff}=T+T_0,
\label{teff}
\end{equation}
where $T_0$ is a constant chosen such that Eq.~(\ref{veff})
reproduces the empirical zero-point energy of a known species.

Now consider the application of this theory to the hydrogen
abstraction reaction, Eq.~(\ref{reaction}).  We will treat
this as a problem in two degrees of freedom $(x,y)$, where
$x$ is the C---H distance for the reacting hydrogen and $y$ is
the H---H distance for the reacting hydrogen.  We will assume
that the C---H---H configuration is linear along the reaction
path.  The angle $\theta$ between the reactive and nonreactive
C---H bonds will be treated as a quadratic polynomial in $x$
that interpolates between 109.5$^\circ$ for CH$_4$,
120$^\circ$ for CH$_3$, and the transition-state geometry
of 102.4$^\circ$ at $x=1.08$ \AA \cite{schatz}.  Then,
\begin{equation}
V_{\rm eff}(x,y)=
{1\over 2\pi\sigma_x\sigma_y}
\int dz_y\, e^{-z_y^2/2\sigma_y^2}
\!\int dz_x\, e^{-z_x^2/2\sigma_x^2}
V(x+z_x,y+z_y),
\label{fullveff}
\end{equation}
where
\begin{equation}
\sigma_x^2={\hbar^2\over 12\mu_x k_B(T_{0,x}+T)},\quad
\sigma_y^2={\hbar^2\over 12\mu_y k_B(T_{0,y}+T)}.
\label{sigmas}
\end{equation}
Note that the mass in Eq. (\ref{veff}) has been replaced by
the reduced mass $\mu_x=m_{\rm H}(1+m_{\rm H}/m_{\rm C})^{-1}$
or $\mu_y=m_H/2$.

For $V(x,y)$ we use Brenner's potential I \cite{brenner},
which is a sum of two-body interactions
\begin{equation}
V(x,y)=V_{\rm CH}(x,y)+V_{\rm HH}(y,x).
\end{equation}
The $V_{\rm CH}$ and $V_{\rm HH}$ each have the form
\begin{equation}
V_i(r,q)=f_i(r)\bigg\{
D^{(R)}_i\exp\Big[ -\beta^{(R)}_i\, \Big(r-r_i^{(e)}\Big)\Big]
-B_i(r,q)D^{(A)}_i
\exp\Big[ -\beta^{(A)}_i\,\Big(r-r_i^{(e)}\Big)\Big]
\bigg\},
\label{vi}
\end{equation}
where $r$ is the coordinate of the primary, two-body,
interaction while $q$ is the coordinate of the ``environment.''
Equation~(\ref{vi}) has the general form of a Morse potential
but is modified by the functions $f_i$ and $B_i$.
$f_i(r)$ is a cutoff function that smoothly interpolates to
zero in the limit of large $r$.  $B_i(r,q)$ is a rather
complicated function that models the effects of nearby atoms
on the primary interaction.  For given $(x,y)$
Eq.~(\ref{fullveff}) can be evaluated by numerical quadrature.
The values
\begin{equation}
T_{0,x}=92 K,\qquad T_{0,y}=582 K
\label{tzeros}
\end{equation}
for the temperature shift parameters
give agreement with the spectroscopically determined
zero-point energies for CH$_4$ \cite{zpech4} and H$_2$
\cite{zpeh2}, respectively.

There is a minor inconsistency in using this procedure
with the Brenner potential.  For the C---H well depth
Brenner simply used the bond energy of the CH molecule
without subtracting the zero-point energy.  Thus, we are
in a sense adding zero-point energy to a potential that
already includes it.  This will make the calculated activation
energy smaller than it ought to be.  We will assume that any
errors from this procedure will be insignificant, since the
quantum effects for C---H interactions are much smaller
than those for H---H interactions.  However, in
principle, one ought to refit the potential using the correct
well depth.  For the H---H potential, which causes most
of the quantum effects, Brenner did use the correct well
depth for H$_2$.

We find that it is possible to accurately approximate
$V_{\rm eff}$ with an analytic expression.  Let
$V_{\rm eff}=\bar{V}_{\rm CH}+\bar{V}_{\rm HH}$, where
\begin{equation}
\bar{V}_i(r,q)=
{1\over 2\pi\sigma_r\sigma_q}
\int dz_r\, e^{-z_r^2/2\sigma_r^2}\!\int dz_q\,
e^{-z_q^2/2\sigma_q^2}\, V_i(q+z_q,r+z_r).
\label{vbar}
\end{equation}
Note that Eq.~(\ref{vbar}) can be written as
\begin{equation}
\bar{V}_i=\bar{V}_i^{(0)}+\bar{V}_i^{(1)}
+\bar{V}_i^{(2)}+\bar{V}_i^{(3)},
\label{vbarsum}
\end{equation}
where
\begin{eqnarray}
&&\bar{V}_i^{(0)}(r,q)=
{1\over \sqrt{2\pi}\sigma_r}
f_i(r) \bigg[
D^{(R)}_i
\int dz\, e^{-\beta^{(R)}\big( r+z-r^{(e)}_i\big)
   -z^2/2\sigma_r^2}
\nonumber\\
&&\qquad\qquad\qquad\qquad\quad
-D^{(A)}_iB_i(r,q)
\int dz\,
e^{ -\beta^{(A)}\big( r+z-r^{(e)}_i\big)-z^2/2\sigma_r^2}
\bigg],
\label{v0hh}
\\
&&\bar{V}_i^{(1)}(r,q)=
{1\over \sqrt{2\pi}\sigma_r}
\int dz\, e^{-z^2/2\sigma_r^2}
\,\big[ f_i(r+z)-f_i(r)\big]
\nonumber\\
&&\qquad\qquad\qquad\qquad\quad \times
\left[ D_i^{(R)}\,
e^{-\beta^{(R)}\, \big( r+z-r^{(e)}_i\big)}
-D_i^{(A)}\,B_i(r,q)
e^{-\beta^{(A)}\, \big( r+z-r^{(e)}_i\big)}\right],
\label{v1hh}
\\
&&\bar{V}_i^{(2)}(r,q)=
-{1\over 2\pi\sigma_q\sigma_r} D_i^{(A)}
\int dz_r\, e^{-z_r^2/2\sigma_r^2}
\int dz_q\, e^{-z_q^2/2\sigma_q^2}
\nonumber\\
&&\qquad\qquad\qquad\qquad\quad \times
\left[ B_i(r,q+z_q)-B_i(r,q)\right]
\, f_i(r+z_r)\,
e^{-\beta^{(A)}\big( r+z_r-r^{(e)}_i\big)},
\label{v2hh}
\\
&&\bar{V}_i^{(3)}(r,q)=
-{1\over 2\pi\sigma_q\sigma_r} D_i^{(A)}
\int dz_r\, e^{-z_r^2/2\sigma_r^2}
\int dz_q\, e^{-z_q^2/2\sigma_q^2}
\nonumber\\
&&\qquad\qquad\qquad\quad \times
\left[ B_i(r+z_r,q+z_q)-B_i(r,q+z_q)\right]
\, f_i(r+z_r)\,
e^{-\beta^{(A)}\big( r+z_r-r^{(e)}_i\big)}.
\label{v3hh}
\end{eqnarray}
The term $\bar{V}_i^{(0)}$ simply ignores the coordinate dependence
of $f_i$ and $B_i$ for purposes of evaluating the integral.
The integrals in Eq.~(\ref{v0hh}) can be evaluated exactly,
giving an expression identical to that for $V_i$ in
Eq.~(\ref{vi}) except with the prefactors $D_i$ replaced by
the temperature-dependent parameters
\begin{equation}
\bar{D}_i^{(R)}=
D_i^{(R)}\, \exp\left[ \Big(\beta^{(R)}\sigma_r\Big)^2/2\right],
\quad
\bar{D}_i^{(A)}=
D_i^{(A)}\, \exp\left[ \Big(\beta^{(A)}\sigma_r\Big)^2/2\right].
\label{dbars}
\end{equation}

For the C---H interaction, we will use
$\bar{V}_{\rm CH}\approx\bar{V}_{\rm CH}^{(0)}$ and ignore the
three correction terms.  For the H---H interaction, since
$\sigma_y^2$ is over twice as large as $\sigma_x^2$, we will
include all of the correction terms but replace them with
approximate analytical expressions.  Let
\begin{equation}
\bar{V}_{\rm HH}^{(1)}=
\phi_R^{(1)}(y)+B_{\rm HH}(y,x)\phi_A^{(1)}(y),
\end{equation}
where
\begin{eqnarray}
&&\phi_R^{(1)}(y)=D^{(R)}_{\rm HH}
\int dz\, \left[ f_{\rm HH}(y+z)-f_{\rm HH}(y)\right]
\exp\left[-\beta^{(R)}\Big(y-r^{(e)}_{\rm HH}+z\Big)
          -z^2/2\sigma_y\right],
\label{phi1R}
\\
&&\phi_A^{(1)}(y)=-D^{(A)}_{\rm HH}
\int dz\, \left[ f_{\rm HH}(y+z)-f_{\rm HH}(y)\right]
\exp\left[-\beta^{(A)}\Big(y-r^{(e)}_{\rm HH}+z\Big)
          -z^2/2\sigma_y\right].
\label{phi1A}
\end{eqnarray}
We find in practice that the functions $\phi_R^{(1)}$ and
$\phi_A^{(1)}$ can be fit
quite accurately as sums of Gaussians,
\begin{equation}
\phi_\alpha^{(1)}\approx
h_\alpha^{(1,1)}
e^{-\big(y-c_\alpha^{(1,1)}\big)^2/{w_\alpha^{(1,1)}}^2}
+h_\alpha^{(1,2)}
e^{-\big(y-c_\alpha^{(1,2)}\big)^2/{w_\alpha^{(1,2)}}^2},
\end{equation}
with the parameters $h$, $c$ and $w$ given by constants or
by polynomials in $(T+T_{0,y})^{-1}$.  The coefficients of
these polynomials can be obtained
by fitting to a set of exact values of the $\phi_\alpha^{(1)}$
from numerical quadrature of Eqs.~(\ref{phi1R}) and
(\ref{phi1A}).

Another function that can be fit in terms of a sum of Gaussians is
\begin{mathletters}
\begin{eqnarray}
\phi^{(2)}(y,x)&=&
-D_{\rm HH}^{(A)}\,
e^{-\beta^{(A)}\big( y-r^{(e)}_{\rm HH}\big)}
\int dz\,\left[ B_{\rm HH}(y,x+z)-B_{\rm HH}(y,x)\right]
\, e^{-z^2/2\sigma_x}\\
&\approx&
f_{HH}(y)
\left[ h^{(2,1)}
e^{-\big(y-c^{(2,1)}\big)^2/{w^{(2,1)}}^2}
+h^{(2,2)}
e^{-\big(y-c^{(2,2)}\big)^2/{w^{(2,2)}}^2}\right] .
\end{eqnarray}
\end{mathletters}

\noindent
In this case the parameters are polynomials in
$(T+T_{0,y})^{-1}$ and $y$.  The second correction term is
given by
\begin{equation}
\bar{V}_{\rm HH}^{(2)}(y,x)=
\phi^{(2)}(y,x)\,\left[ f_{\rm HH}(y)
e^{{\beta^{(A)}}^2\sigma^2/2}-\phi^{(1)}/D^{(A)}_{\rm HH}
\right].
\end{equation}
Finally, we express the third correction term as
\begin{equation}
\bar{V}^{(3)}(y,x)=
f_{\rm CH}(x)f_{\rm HH}(y)\phi^{(3)}(x,y)/(T+T_{0,y}),
\label{vbar3}
\end{equation}
with $\phi^{(3)}$ fit as a polynomial in $x$ and $y$.

Using polynomials of at most degree 2 for the parameters
we can fit the exact Gaussian averaged potential with
an accuracy of $\pm 0.005$ kcal/mol in the
vicinity of the transition state for $T$ as low as 300 K.
A table of fitting parameters and a full error analysis for the 
the $\phi^{(j)}$ will be presented elsewhere.

\section{Results}

We obtain a temperature-dependent activation energy as the
difference between $V_{\rm eff}$ evaluated at the CH$_5$
saddle point and at the reactants well.  In Fig.~1 the
solid curve shows our quantum mechanical result for $E_a$,
calculated from the Gaussian average with standard deviations
$\sigma_i$ in terms of effective temperatures according
to Eqs.~(\ref{sigmas}).  This result
is in effect an interpolation between the high-temperature
path integral analysis and the empirical low-temperature
limit.  For $T>300$ K we find that $E_a$ can be accurately
fit with a quadratic polynomial in $T^{-1}$,
\begin{equation}
E_a(T)\approx 12.07 {\rm\ kcal\ mol^{-1}}
- (741.2 {\rm\ kcal\ mol^{-1}\ K})\, T^{-1}
+ (7.47\times 10^4 {\rm\ kcal\ mol^{-1}\ K^2})\, T^{-2}.
\label{EaofT}
\end{equation}
The dashed curve shows the result of a purely
high-temperature analysis, with the $\sigma_i$ in terms of
the actual temperature.  The two results are in general
agreement down to approximately 600 K. (The unsteadiness
in the curves at low $T$ is due to uncertainty in
the determination of the geometry of the activated
complex.)

The corresponding Arrhenius curves are shown in Fig. 2, where
they are compared with classical collision theory (the dotted
curve, corresponding to a temperature-independent activation
energy equal to $\lim_{T\to\infty}E_a(T)$\ ), with a recent
multidimensional semiclassical variational TST analysis
\cite{truong} (the dash-dot curve), and with experimentally
determined reaction rates \cite{shaw,smp,rabinowitz}.  Our
theoretical treatment yields the activation energy, but not
the Arrhenius prefactor, which must be fit to the experimental
results.  The theory for the $T$ dependence of $E_a$
is exact in the high-temperature limit.  Therefore, it is
best to consider only high-temperature rate constants for
the purpose of determining the prefactor.  We will assume that
the measurements by Rabinowitz {\it et al.} \cite{rabinowitz},
from the recent Brookhaven flash photolysis-shock tube study,
are the most reliable of the high temperature results.  Fitting
the classical collision theory expression
\begin{equation}
k=A_0 T^{1/2} e^{-E_a/RT}
\label{arrhenius}
\end{equation}
to the the Brookhaven results, with our theoretical formula for
the function $E_a(T)$, gives
$\log_{10}(A_0\,/{\rm\,cm^3s^{-1}})=-11.536$.
It is clear from Fig.~2 that the Gaussian average with $T_{\rm eff}$
is in better agreement with the low temperature measurements than
are the Gaussian average with the actual $T$ or the classical 
theory with $E_a$ independent of temperature.

\section{Discussion}

\subsection{Comparison with transition-state theory}

It can be seen in Fig.~2 that our theory (the Gaussian average with
$T_{\rm eff}$) and Truong's TST \cite{truong} both predict a distinct
upward curvature in the Arrhenius plot at low temperature.  They
disagree at high temperature, with our analysis predicting a
slower reaction rate than that predicted by the TST study.

Our theoretical approach differs from TST in the way in
which we describe the classical dynamics and in the way we
include quantum effects.  In TST the expression for the rate
constant can be
expressed approximately in the form $k=a_0T^x\Gamma e^{-E_a/RT}$,
where $a_0$ and $x$ are temperature-independent constants
and $\Gamma(T)$ is a quantum correction \cite{frost,clark}.
The reactants and the transition state are assumed to be in
equilibrium, and the factor $T^x$ comes from the resulting
equilibrium constant, expressed as a ratio of partition functions.
For CH$_4$+H$\,\rightleftharpoons\,$CH$_5$ the value of $x$ is
3/2.  Thus, we can express the TST result in the form
\begin{equation}
k=a_0T^{3/2}e^{-E_a/RT},
\label{ktst}
\end{equation}
with $E_a$ equal to a constant for classical TST and
$E_a(T)=E_a^{(cl)}\ln\big(\Gamma(T)\big)$ for quantum TST.
Our approach is to use collision theory for the classical
dynamics, which gives $x=1/2$, and to determine $E_a(T)$ directly
from the temperature-dependent effective potential instead
using one of the standard semiclassical expressions for
$\Gamma(T)$.  However, we could just as well have used the
effective-potential values for $E_a$ in Eq.~(\ref{ktst})
instead of using the collision-theory expression
Eq.~(\ref{arrhenius}).

A clearer comparison of theory and experiment is given by
Fig.~3, which shows the difference between $\log_{10}k$ and
$\log_{10}k_{\rm cl}$, where $k_{\rm cl}$ is the rate constant
predicted by classical theory.  We obtain $k_{\rm cl}$ from
Eq.~(\ref{arrhenius}) by setting $E_a=12.07$ kcal/mol, which
is the value given by the Brenner potential.  We use the same
value of $A_0$ as with the $T$-dependent $E_a$, since the quantum
and classical theories have the same high-$T$ limit.
The solid curve in Fig.~3 corresponds to our
quantum collision theory, using Eq.~(\ref{arrhenius}) with
the effective-potential $E_a(T)$ and with
$T_{\rm eff}$ in the standard deviations for the Gaussian averaging.
The dotted curve results from using this same $E_a(T)$
in the TST expression
Eq.~(\ref{ktst}).  The solid curve is clearly in better
agreement with the experimental points.

Figure~3 also shows Truong's
variational TST results \cite{truong} (the dash-dot curve),
which were based on
{\it ab initio} calculations of the potential, and it shows
results from a variational TST study by Joseph {\it et al.}
\cite{joseph} (the dash-dot-dot curve),
which used an analytical semiempirical
potential energy function.  The prefactor $a_0$ used for
our effective-potential TST was chosen to give
agreement with the high-temperature experimental results.  Note,
however, that different values of $a_0$ would lead to qualitative 
agreement between the effective-potential TST and the variational
TST studies.  Although there is a fair
amount of scatter in the experimental results, our 
quantum collision theory and the
TST studies appear to agree equally well with the low-temperature
experimental points, where quantum effects are expected to be most
important.  This is consistent with previous studies for other
systems \cite{doll,lauderdale}, which showed agreement between
quantum effects calculated from semiclassical TST quantum
corrections and those calculated from effective potentials.

At high temperatures the results from Truong's TST results lie
above most of the experimental points.  The results from
semiempirical TST are in better agreement with experiment, but
extrapolation of them to higher $T$ would give a larger
rate constant than does our collision theory.
Since barrier recrossing becomes
more likely at higher $T$, one can expect that TST will
increasingly overestimate the reaction rate as $T$ increases.
Truong used an {\it ab initio} result
for the barrier height (16.3 kcal/mol) that was higher than
the value used by Joseph {\it et al.} (12.9 kcal/mol) or the
value used by us (12.07 kcal/mol).  However, as $T$ increases,
the value used for $E_a$ becomes less significant than the
value used for the exponent $x$.  At high $T$ the assumption
on which collision theory is based, that only the translational
degrees of freedom of the colliding molecules need to be taken
into account, is reasonable since the lifetime of the activated
complex will typically be too short for there to be significant
conversion of translational energy into vibrational or rotational
energy before it breaks apart.  At lower temperature TST will
be more accurate than collision theory, but the error introduced
by using $x=1/2$ will become relatively less important than 
errors in the value of $E_a$.  The excellent agreement with
experiment from our quantum collision theory would seem to
indicate that this is what happens for the particular reaction
considered here.  In any case, the approach we have presented
for calculating quantum effects can be used with TST as
well as with collision theory, and it is conceivable that for
other reactions, at intermediate temperatures, TST might be
the more appropriate choice.

\subsection{Comparison of analytical expressions for the
rate constant}

In Fig.~4 we compare our calculated results with various
``best fits'' of the experimental points
\cite{shaw,smp,baulch,rabinowitz}.  The fitting functions
are given in Table~1.  For Fig.~4 we computed our rate constants
using analytic expresions for $E_a(T)$.  The solid curve
was computed using the quadratic fitting function in
Eq.~(\ref{EaofT}).  This curve is almost indistinguishable from
the corresponding curve in Fig.~3, which was computed from 
our actual values for the $E_a$.  The dashed curve
results from a linear fit for $E_a(T)$, which yields an
expression for $k$ that has only 3 parameters.  Most of
the empirical fits in Fig.~4 are also 3-parameter expressions.

Our results in Fig.~4 are in better agreement with the
experimental points than are any of the empirical expressions.
The expressions of Shaw \cite{shaw} and of Baulch {\it et al.}
\cite{baulch} (the dash-dot and dash-dot-dot curves, respectively)
seem to be too low in the low-$T$ region.  Both
of these fits were carried out before the data from the
high-$T$ measurements by Rabinowitz {\it et al.} \cite{rabinowitz}
were available.  Shaw's fit agrees with these new data while
the fit by Baulch {\it et al.} appears to be too high in
the high-$T$ region.  Not surprisingly, the fit by Rabinowitz
{\it et al.} (the dash-dot-dot-dot curve) seems accurate
at high-$T$.  However, it appears
to be too high at low $T$.  Note that only one of our
parameters (the prefactor $A_0$) is fit to the experimental
results.  The remaining parameters are calculated
theoretically.  (Of course, other experimental results, such
as the bond dissociation energies of CH$_4$ and H$_2$, enter
our theory indirectly since we based our calculations on an
empirical potential function \cite{brenner}.)

The most significant difference between our expressions for
$k$ and the various fits is the behavior that is
predicted at very high $T$, above 2000 K.  Our expressions
predict a significantly lower reaction rate in this region
than do any of the others.
Unfortunately, there are no available experimental results
above 2000 K with which to compare.  The quantum effects at very
high $T$ are insignificant.  Therefore, the accuracy of our
results in that region depends only on the accuracy
of the activation energy given by the Brenner potential and
on the accuracy of the Arrhenius expression from classical
collision theory.

\subsection{Conclusions}

We have shown that the Arrhenius expression from classical
collision theory with $E_a$ expressed as a polynomial in
$1/T$ gives agreement with the experimental
measurements for the rate constant for CH$_4$+H that is
at least as good as that from
fits that include $T$ dependence in the prefactor.  This
is especially striking since only one of the parameters in
our expressions is determined by fitting to
the experimental points.  The fact that the other parameters
are determined theoretically suggests that our expressions
are more likely to be dependable at temperatures above
2000 K or below 350 K, where experimental results are
unavailable.

These calculations require that the potential energy surface
be known.  Otherwise, there is no choice but to choose a
functional form for $k(T)$ containing parameters that are all
determined by fitting to measured reaction rates.  Traditionally,
an Arrhenius expression is used with the temperature dependence
in the prefactor.  However, the theory developed here provides
a justification for choosing a functional form with temperature
dependence in the activation energy.

The success of our theory at describing this reaction rate
over a wide temperature range has two important implications
for molecular dynamics
simulations of chemical reactions.  First, it supports the
use \cite{garrison} of the Brenner hydrocarbon potential
in those studies.  This is an empirical potential,
parameterized to a data set consisting of properties of
{\it stable} chemical species, yet it seems to be able to
accurately model the unstable CH$_5$ transition state, since the
curvature of the Arrhenius plot depends sensitively on the
topography of the potential in the vicinity of the saddle
point.

The second implication is that it is possible to use a
classical molecular dynamics computation, with a
Gaussian-averaged effective potential, to model processes in
which quantum effects are important.  The evaluation of the
potential function is the most time-consuming step in these
computations.  Therefore, the use of numerical quadrature
to perform the Gaussian averaging would be impractical.
However, the use of our {\it analytic} approximation, given by
Eqs.~(\ref{vbarsum}) through (\ref{vbar3}), will have only a
minor effect on the computational cost.

\acknowledgments{This work was supported by the Robert A.
Welch Foundation and the National Science Foundation.}

{\tighten
\widetext
\begin{table}
\caption{Expressions for the rate constant
for CH$_4$+H$\,\to\,$CH$_3$+H$_2$.}
\begin{tabular}{lll}
rate constant, cm$^3$ molecule$^{-1}$ s$^{-1}$
&method&reference\\
\tableline
\\
$k=3.02\times10^{-10}\,\exp(-6631/T)$
&empirical fit&Sepehrad\\
&&{\it et al.}\protect\tablenotemark[1]\\
\\
$k=2.35\times10^{-17}\,T^2\exp(-4449/T)$
&empirical fit&Shaw\protect\tablenotemark[2]\\
\\
$k=2.18\times10^{-20}\,T^3\exp(-4045/T)$
&empirical fit&Baulch\\
&&{\it et al.}\protect\tablenotemark[3]\\
\\
$k=6.4\times10^{-18}\,T^{2.11}\exp(-3900/T)$
&empirical fit&Rabinowitz\\
&&{\it et al.}\protect\tablenotemark[4]\\
\\
$k=2.63\times10^{-12}\,T^{1/2}
 \exp\big(-6076/T$
&collision theory with
&this study\protect\tablenotemark[5]\\
\qquad\quad\quad\quad $+3.730\times10^5/T^2-3.76\times10^7/T^3\big)$
&effective potential;&\\
&quadratic expression for $E_a$&\\
\\
$k=2.63\times10^{-12}\,T^{1/2}
 \exp\big(-6076/T$
&collision theory with
&this study\protect\tablenotemark[4]\\
\qquad\qquad\qquad\qquad\quad\quad\quad\quad\, $+2.759\times10^5/T^2\big)$
&effective potential;&\\
&linear expression for $E_a$&\\

\end{tabular}
\tablenotetext[1]{Ref. \cite{smp}.}
\tablenotetext[2]{Ref. \cite{shaw}.}
\tablenotetext[3]{Ref. \cite{baulch}.}
\tablenotetext[4]{Ref. \cite{rabinowitz}.}
\tablenotetext[5]{From Gaussian average of potential with $T_{\rm eff}$
in the standard deviations.}
\end{table}}

\narrowtext

\begin{figure}
\caption{Temperature dependence of the activation energy.
The dotted line shows the classical,
temperature-independent, result.  The dashed curve was
calculated from the Gaussian-averaged potential using
the standard deviations given by Eq.~(\protect\ref{sigma}).
The solid curve was calculated from the Gaussian-averaged
potential with $T$ in Eq.~(\protect\ref{sigma}) replaced
by $T_{\rm eff}$ according to Eqs.~(\protect\ref{sigmas})
and (\protect\ref{tzeros}).}
\end{figure}

\begin{figure}
\caption{Arrhenius plot of rate constant vs. 1000 K/$T$
for the reaction CH$_4$+H$\,\to\,$CH$_3$+H$_2$, with $k$
in units of cm$^3$ molecule$^{-1}$ s$^{-1}$.  Experimental
measurements, from Refs.~\protect\cite{shaw},
\protect\cite{smp}, and \protect\cite{rabinowitz}, are
indicated by dots.  The dotted line
corresponds to classical collision theory, while the dashed
and solid curves correspond to our quantum theory using $T$
and $T_{\rm eff}$, respectively, in the standard deviations.
The dash-dot curve shows results from Truong's variational TST
\protect\cite{truong}.}
\end{figure}

\begin{figure}
\caption{Difference between $\log_{10}k$ and $\log_{10}k_{\rm cl}$
vs. 1000 K/$T$, where $k_{\rm cl}$ is the rate constant from
classical collision theory with temperature-independent $E_a$.
The units of $k$ are \hbox{cm$^3$ molecule$^{-1}$ s$^{-1}$}.
The symbols indicate experimental points from the following
references: Rabinowitz {\it et al.} \protect\cite{rabinowitz}
(+); Sepehrad {\it et al.} \protect\cite{smp}
($\Box$); and the various studies reviewed by Shaw
\protect\cite{shaw} ($\diamond$).  The dashed
and solid curve shows results from the present study,
using $T_{\rm eff}$ in the standard
deviations of the Gaussian average.
The dotted curve corresponds to classical TST, according
to Eq.~(\protect\ref{ktst}), with $E_a(T)$ from
the Gaussian average.
The dash-dot curve corresponds to Truong's variational TST
\protect\cite{truong}, which uses {\it ab initio}
calculations for the potential energy.
The dash-dot-dot curve corresponds to the variational TST
of Joseph {\it et al.} \protect\cite{joseph}, which uses
a semiempirical potential function (``J3'').}
\end{figure}

\begin{figure}
\caption{Difference between $\log_{10}k$ and $\log_{10}k_{\rm cl}$
vs. 1000 K/$T$, using the analytical expressions for $k(T)$
in Table~1.  $k_{\rm cl}$ is the rate constant from
classical collision theory with temperature-independent $E_a$.
The units of $k$ are \hbox{cm$^3$ molecule$^{-1}$ s$^{-1}$}.
The symbols indicate experimental points, labeled as in
Fig.~3.  The dotted, dash-dot, dash-dot-dot, and
dash-dot-dot-dot curves are
the empirical fits from Sepehrad {\it et al.} \protect\cite{smp},
Shaw \protect\cite{shaw},
Baulch {\it et al.} \protect\cite{baulch}, and
Rabinowitz {\it et al.} \protect\cite{rabinowitz},
respectively.  The solid and dashed curves are from the
present study with a quadratic and a linear fit, respectively,
for $E_a$.}
\end{figure}

\end{document}